\documentclass[12pt,a4paper]{article}

\usepackage[utf8]{inputenc}
\usepackage[english]{babel}
\usepackage{amsmath}
\usepackage{amsfonts}
\usepackage{amssymb}
\usepackage{csquotes}
\usepackage{graphicx}
\usepackage{geometry}
\usepackage{setspace}
\usepackage{xcolor}
\usepackage{bbm}
\usepackage{subcaption}
\usepackage{float}
\usepackage{siunitx}
\DeclareSIUnit{\litre}{l}
\DeclareSIUnit{\Molar}{M}
\usepackage{booktabs}
\usepackage{makecell}

\usepackage{setspace} 
 
\usepackage{lineno} 

\usepackage{upgreek,textgreek}

\usepackage[square,numbers,sort&compress]{natbib}

\definecolor{ulmgruen}{rgb}{0.3272,0.6667,0.1098}

\newcommand{\N}{\mathbb{N}}
\newcommand{\W}{W}

\newcommand{\STEM}{I}
\newcommand{\phaseWise}{I_\text{phase}}
\newcommand{\peptideWise}{I_\text{peptide}}
\newcommand{\virionWise}{I_\text{virion}}

\newcommand{\cellPhase}{\mathcal{C}}
\newcommand{\peptidePhase}{\mathcal{P}}
\newcommand{\peptideSet}{P}
\newcommand{\peptide}{p}
\newcommand{\virionSet}{V}
\newcommand{\virion}{v}

\newcommand{\ved}{\text{diam}}
\newcommand{\volume}{\text{vol}}
\newcommand{\ssa}{\sigma_\text{ssa}}

\newcommand{\surfArea}{A}
\newcommand{\density}{\rho}
\newcommand{\coverageRatio}{\theta_\text{cov}}
\newcommand{\minDist}{\text{dist}_\text{min}}

\date{\vspace{-4ex}}
\title{Statistical analysis of virion-cell interactions mediated by peptide nanofibrils and peptide amphiphiles using STEM tomography}
\author{\parbox{\textwidth}{\centering
  Philipp Rieder$^{1,*,\dagger}$, 
  Julia La Roche$^{2,*,\dagger}$, 
  Orkun Furat$^{1,3}$, 
  Annalena Kuhn$^4$,
  Lena Rauch-Wirth$^4$,
  Kübra Kaygisiz$^{5}$, 
  Fabian Zech$^4$, 
  Jan M\"unch$^4$, 
  Clarissa Read$^2$, 
  R\"udiger Gro\ss{}$^{4,*}$, 
  Volker Schmidt$^1$
}}

\begin{document}
\maketitle
\footnotesize
\noindent$^1$Institute of Stochastics, Ulm University, D-89069~Ulm, Germany\\
$^2$Central Facility for Electron Microscopy, Ulm University, D-89081~Ulm, Germany\\
$^3$SDU Applied AI and Data Science Unit, University of Southern Denmark, DK-5230 Odense, Denmark\\
$^4$Institute of Molecular Virology, Ulm University Medical Center, D-89081~Ulm, Germany\\
$^5$Department of Chemistry, Massachusetts Institute of Technology, Cambridge, Massachusetts,\\\phantom{$^*$} 02139 United States\\
$^*$Corresponding authors:\\ 
\phantom{$^*$}philipp.rieder@uni-ulm.de (P.R.),
julia.la-roche@uni-ulm.de (J.L.), 
ruediger.gross@uni-ulm.de (R.G.)\\
$^\dagger$P.R. and J.L. contributed equally to this paper.\\[1em]
\normalsize



\section*{Abstract}

Peptide nanofibrils (PNFs) and peptide amphiphiles (PAs) are promising tools for enhancing viral transduction and gene transfer.
However, quantitative insight into how their supramolecular architecture governs virion-cell interactions is limited.
Here, we introduce a framework for the acquisition, processing, and statistical analysis of scanning transmission electron microscopy (STEM) tomograms to objectively quantify peptide-virion-cell interactions.
Using four transduction-enhancing peptides (D4, Vectofusin-1, palmitic acid-PA (pal-PA), and eicosapentaenoic-PA (eic-PA)), peptide aggregate morphology, interfacial contact areas, and the spatial organization of virions with respect to peptides and cells were analyzed using advanced geometric descriptors.
All peptides efficiently captured virions, resulting in few free virions, but they differ in how strictly virions were spatially confined near the cell surface. 
These differences reflect alternative spatial organization strategies, which are likely  crucial factors influencing transduction-enhancing efficacy.
Our approach provides a novel, generalizable method to evaluate infection-enhancing nanomaterials and guides the rational design of next-generation peptide assemblies for therapeutic viral delivery.\\

\section*{Graphical abstract}
\begin{figure}[h!]
    \centering
    \includegraphics[width=1.1\linewidth]{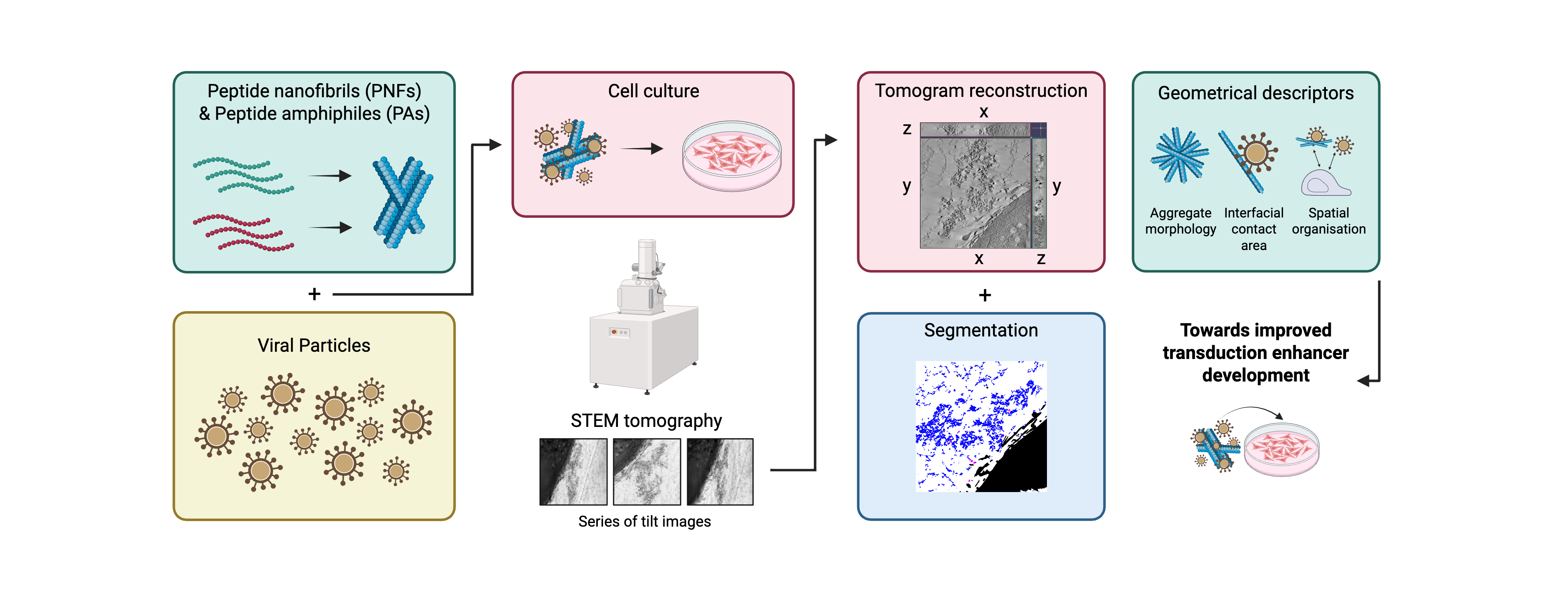}
    \label{fig:placeholder}
\end{figure}


\newpage
\section{Introduction}
Peptide nanofibrils (PNFs) have emerged as modulators of viral transduction (delivery of genetic material) by promoting the attachment of virions to target cells and facilitating their uptake through physicochemical interactions. 
This phenomenon was first identified in human semen, where amyloidogenic peptides derived from abundant proteins such as prostatic acid phosphatase (PAP)~\cite{Muench2007} or semenogelins~\cite{Roan2014} assemble into fibrillar structures that enhance HIV-1 infection. 
Importantly, HIV-1 serves as a prototypical system for retroviral and lentiviral vectors widely used in gene therapy, as these vectors retain key structural and entry-related features of the native virus. Consequently, mechanistic insights into PNF-mediated enhancement of HIV-1 infection are directly relevant for improving retroviral transduction strategies. However, such enhancement effects are not universal across all enveloped viruses or fibrillar systems and depend critically on peptide composition, supramolecular organization, and the specific interaction context.

A prominent application of retroviral gene transfer is chimeric antigen receptor (CAR) T cell therapy, where patient-derived T cells are genetically engineered using predominantly lentiviral or $\gamma$-retroviral vectors to express tumor-targeting receptors. 
Despite its clinical success in treating hematological malignancies, the efficiency of this viral transduction step remains a major bottleneck and contributes substantially to the high cost of these therapies~\cite{Patel2025}. 
PNFs offer a promising strategy to overcome this limitation, as they can markedly increase transduction efficiencies~\cite{RauchWirth2025}. 
Mechanistically, this enhancement is primarily attributed to electrostatic interactions between positively charged fibrils and negatively charged viral particles, leading to the formation of supramolecular assemblies that promote virus–cell contact and uptake.

Mechanistically, transduction enhancement by peptide assemblies is primarily driven by electrostatic interactions that increase virion attachment to the cell surface~\cite{Roan2009}. 
Positively charged fibrils bind negatively charged viral particles, forming supramolecular complexes that concentrate virions at the plasma membrane and along cellular protrusions such as filopodia. 
This increased local enrichment facilitates viral entry, which can occur either directly via membrane fusion at the cell surface or via endocytic uptake, depending on the viral system and cellular context~\cite{RauchWirth2025}.
From a biophysical perspective, this process involves several coupled steps: (i) self-assembly of peptide monomers into fibrils and higher-order aggregates, (ii) binding and spatial organization of virions within these assemblies, and (iii) interaction of peptide–virus complexes with the cell surface, ultimately enabling entry into the cell.

The two peptides Vectofusin-1 (Vect-1) and D4 are representative examples of peptide-based transduction enhancers that rely on self-assembly into PNFs but differ significantly in structure and peptide aggregation behavior. Vect-1 is a cationic 26-mer peptide (KKALLHAALAHLLALAHHLLALLKKA-NH\textsubscript{2}) that forms $\alpha$-helical PNFs and is currently being explored for use in CAR-T cell manufacturing~\cite{radek_vectofusin-1_2019, vermeer_vectofusin-1_2017, irving_choosing_2021}. In contrast, the shorter 8-mer peptide D4 (RRIFIISM) was developed by \textit{in silico} screening and structure-activity relationship analysis and assembles into chemically stable, $\beta$-sheet-rich PNFs that form small, biodegradable aggregates. 
D4 exhibits superior transduction efficiency in primary T and NK cells, which has been attributed to its high positive net charge and favorable aggregation properties that enhance viral binding while minimizing cytotoxicity and immunogenicity~\cite{rauch-wirth_optimized_2023, RauchWirth2025,LaRoche2026D4}. In addition to PNFs, peptide amphiphiles (PAs) constitute a distinct class of transduction enhancers composed of a peptide sequence linked to an aliphatic chain that assembles into various nanostructures~\cite{hartgerink_self-assembly_2001,cui_selfassembly_2010}. Palmitoyl-VVVAAAKKK-NH\textsubscript{2} (pal-PA) enhances retroviral transduction through nanofibril formation, similar to other fibril-based enhancers. By contrast, eicosapentaenoyl-VVVAAAKKK-NH\textsubscript{2} (eic-PA) achieves a comparable enhancement of viral delivery without forming fibrils but amorphous structures and displays exceptionally high cellular degradability (99\%), a good prerequisite for clinical use~\cite{Kaygisiz2024, LaRoche2025}.

Electron microscopy provides direct ultrastructural insight into these processes and is a key tool for the rational optimization of peptide-based transduction enhancers. 
Systematic analysis of electron tomography datasets has been limited by the labor-intensive nature of manual segmentation and evaluation. 
Methods from classical computer vision have been employed to accelerate the segmentation of specific features in tomograms, e.g.~\cite{weber2021multidimensional,BADIAMARTINEZ2012Three}.
In recent years, improved artificial intelligence (AI)-assisted segmentation tools have become more popular~\cite{Ronneberger2015UNEt,Weber2023segmentation,heebner2022deep,dragonfly}.

Such AI-assisted image-processing approaches enable the segmentation of a large number of 3D image datasets.
However, data at this scale cannot be reliably analyzed by visual inspection alone.
To ensure a robust analysis, it is necessary to establish mathematical frameworks for their statistical evaluation.
These frameworks are based on geometric descriptors that condense specific features of the complex voxelized 3D morphologies into scalar-valued quantities.
Such descriptors range from simple measures, such as the diameter of virus-like particles~\cite{martin2016distinct,Maldonado2016DistinctMorphology}, to more sophisticated characteristics, including aspect ratio and fibril width~\cite{McCraw2018structural,Weber2023segmentation}.
Similar approaches have also proven promising for investigating interactions between different phases in technical materials~\cite{RIEDER2025115602}.
By applying mathematical concepts such as (multivariate) kernel density estimation, correlation structures among these descriptors, and consequently among properties of the imaged structures can be revealed, thereby contributing to their understanding.

In this study, we introduce a framework to statistically analyze peptide-virion-cell interaction based on three-dimensional datasets generated by scanning transmission electron microscopy (STEM) tomography. 
The tomograms are segmented phase-wise and instance-wise utilizing AI-assisted approaches as well as methods from classical image processing. Based on these segmentations, the computation of several geometric descriptors is performed. The statistical analysis of these descriptors allows for a quantitative comparison of different PNFs and PAs, providing mechanistic insight into their function as viral transduction enhancers.

\section{Materials and Methods}

\subsection{Sample preparation}
\subsubsection{Peptide nanofibrils}
The peptide D4 (RRIFIISM) was custom-produced by GaloreTx Pharmaceuticals (Udupi, India), Vectofusin-1 (Vect-1, KKALLHAALAHLLALAHHLLALLKKA-NH\textsubscript{2}) was purchased from Miltenyi Biotec (130-111-163), and eicosapentaenoic peptide amphiphile (eic-PA, eicosapentaenoic acid-VVVAAAKKK‐NH$_2$) from Bachem AG. Pal-PA (palmitic acid-VVVAAAKKK‐NH$_2$) was synthesized using an automated microwave peptide synthesizer (CEM, Liberty BlueTM) according to our established protocol~\cite{Kaygisiz2024}.
Note that in the referenced study, pal-PA was termed PA1 and eic-PA Pat10. Fatty acid conjugation was performed according to previously reported methods~\cite{Pashuck2010, Paramonov2006}. 
It is important to note that  D4 is capable of forming fibrils with lengths on the nanometer and micrometer scale. As micrometer-scale fibrils rarely occur when cells are in close proximity, they were excluded from the present study.

\subsubsection{Cells}
HeLa cells were obtained from ATCC (CRM-CCL-2) and were cultured in Dulbecco's Modified Eagle Medium (DMEM, Gibco, 11965092), supplemented with 
\SI{100}{U/\milli\liter} penicillin, 
\SI{2}{\milli\Molar} L-glutamine (PAN-Biotech, P04-80050),
\SI{100}{\micro\gram/\milli\liter} streptomycin (PAN-Biotech, P06-07050), 
and 10\% inactivated fetal bovine serum (FBS, Gibco, 10437028).

\subsubsection{Virion stock solutions}
The non-infectious virions (murine leukemia virus-Gag yellow fluorescence virus-like particles) were produced as reported in our previous study~\cite{LaRoche2026D4}. For their use in electron microscopy, the virion-containing supernatant was sterile filtered (\SI{0.45}{\micro\meter}) and purified by ultracentrifugation and a sucrose gradient (20\%-60\%).

\subsubsection{Scanning transmission electron microscopy}\label{ssec:STEM}
The interaction between peptide nanofibrils, virions, and HeLa cells was analyzed by performing STEM tomography as described in~\cite{LaRoche2025,LaRoche2026D4, wieland_scanning_2025}. Therefore, 50,000 HeLa cells were seeded on glow-discharged carbon-coated sapphire disks (\SI{3}{\milli\meter} diameter, \SI{50}{\micro\meter} thickness; Wohlwend GmbH) in 12-well plates. 
The following day, peptide nanofibrils (\SI{2}{\micro\gram/\milli\liter}) were pre-incubated with virions ($1:10$ dilution) at room temperature for \SI{10}{\minute}, after which the mixture was added to the cells and incubated for \SI{1}{\hour} at \SI{37}{\celsius}.
Following incubation, the samples were processed according to our established protocols~\cite{wieland_scanning_2025, walther_freeze_2002,LaRoche2026D4}. Therefore, samples were cryo-immobilized by high-pressure freezing using a HPF Compact 01 system (Wohlwend GmbH), processed by freeze substitution in an EM AFS2 unit (Leica), and embedded in epoxy resin (Embed-812, Science Services). 
The resin blocks were sectioned into approximately \SI{800}{\nano\meter} thick sections, which were mounted onto 200-mesh copper grids with parallel bars (Plano GmbH). 
Before tomogram acquisition, sections were treated with \SI{25}{\nano\meter} colloidal gold particles (AURION Immuno Gold Reagents \& Accessories) serving as fiducial markers to facilitate accurate alignment of tilt images during tomographic reconstruction.

\begin{figure}[H]
    \centering
    \includegraphics[width=.74\textwidth]{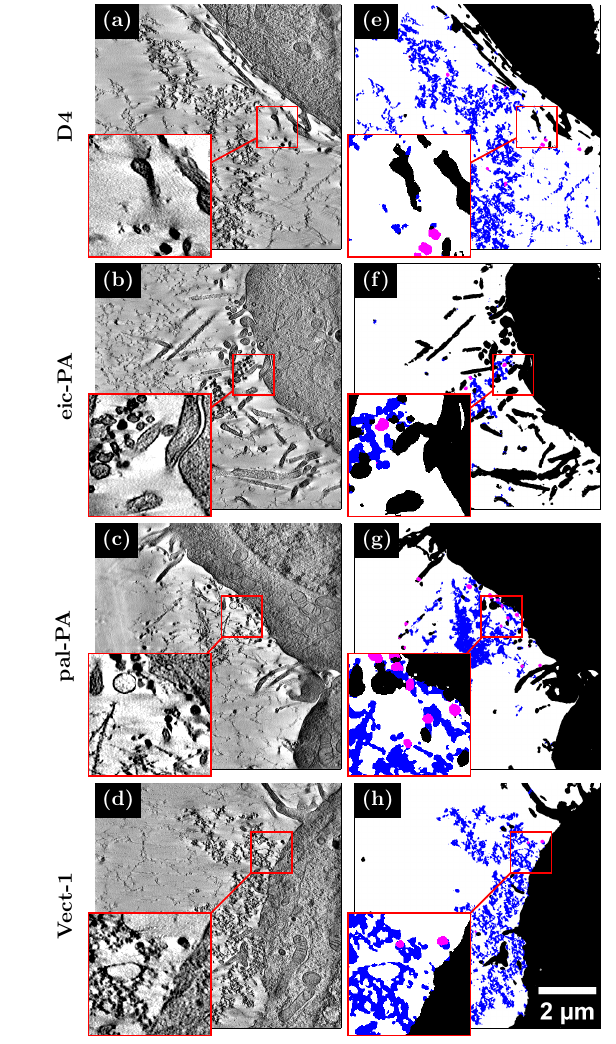}
    \caption{Exemplary cross sections of the investigated peptides D4, eic-PA, pal-PA and Vect-1.
    (a)-(d) show virtual cross sections of the tomograms whereas (e)-(h) display the corresponding phase-wise segmentation, see Section~\ref{sec:segmentation} for details. 
    Cells are indicated in black, peptides in blue and virions in magenta.
    }
    \label{fig:peptidesOverview}
\end{figure}

Tomographic data were acquired using a JEM-2100F field emission transmission electron microscope (JEOL GmbH) operated in STEM mode at an accelerating voltage of \SI{200}{\kilo\volt}. 
The tilt series was collected over an angular range from \SI{-72}{\degree} to \SI{+72}{\degree}  in \SI{1.5}{\degree} increments, resulting in a total of 97 projection images. 
Tilt image alignment was carried out using the IMOD software package, and three-dimensional reconstructions were generated by weighted back-projection, as previously described~\cite{wieland_scanning_2025}. 
The resulting datasets have a voxel size of $\SI{8.3}{\nano\meter}$. 
Lastly, the grayscale values of the tomograms were scaled so that the entire range of 8bit integers is used. 
In total, 17 tomograms were acquired, three corresponding to D4, six to eic-PA, four to pal-PA, and four to Vect-1.
Exemplary cross sections of tomograms showing the four different peptides with virions on cells are presented in Figures~\ref{fig:peptidesOverview}(a)-(d).

\subsubsection{Lentiviral transduction enhancement}
For the biological confirmation of the results obtained from the statistical analysis based on the STEM images, 
R5-tropic HIV-1 NL4-3 infection of TZM-bl reporter cells was used as a quantitative measurement for the transduction enhancing capabilities of peptide nanofibrils~\cite{Muench2007}. Briefly, 10,000 TZM-bl cells were seeded one day prior, virus stock solutions were mixed with peptides, resulting in concentrations up to \SI{12.5}{\micro\gram/\milli\liter} (on cells), and incubated for 10 min at room temperature. After the addition of the mixtures to the cells and incubation for 2 days, transduction rates were measured by $\beta$-galactosidase assay.

\subsection{Image segmentation}\label{sec:segmentation}
To enable a quantitative comparison between the different peptides and their interactions with  virions and cells in Section~\ref{sec:Discussion}, both phase-wise  and instance-wise segmentations of the tomograms are determined.
Note that for both the peptide phase and the virion phase, a separate instance-wise segmentation is determined, referred to as aggregate-wise and virion-wise segmentation, respectively.
For that, let $\STEM\colon\W\rightarrow\{0,1,\ldots,255\}$ denote a grayscale image derived from STEM tomography measurements as described in Section~\ref{ssec:STEM}.
Here $\W=\{1,\ldots,w_1\}\times\{1,\ldots,w_2\}\times\{1,\ldots,w_3\}\subset\N^3$ denotes the discretized, cuboidal observation window of $\STEM$, where $\N=\{1,2,3,\ldots\}$ represents the set of natural numbers and $w_1=w_2=1024$.
The thickness $w_3\in\{65,\ldots,146\}$ varies for each tomogram $\STEM$ since they were cropped to ensure that they contain no imaging artifacts.
First, the phase-wise segmentation of $\STEM$ is defined as mapping $\phaseWise\colon\W\rightarrow\{0,1,2,3\}$, given by
\begin{align*}
    \phaseWise(x)=
    \begin{cases}
        1, \quad &\text{if $x$ corresponds to the cell phase,}\\
        2, \quad &\text{if $x$ corresponds to the virion phase,}\\
        3, \quad &\text{if $x$ corresponds to the peptide phase,}\\
        0, &\text{else,}
    \end{cases}
\end{align*}
which assigns each $x\in\W$ to the cell, virion, peptide, or background phase.
The cell phase is denoted by $\cellPhase=\{x\in\W\colon \phaseWise(x)=1\}$ and the peptide phase by $\peptidePhase=\{x\in\W\colon\phaseWise(x)=3\}$. 
In practice, the phase-wise segmentation was derived following an approach similar to that presented in~\cite{heebner2022deep}, utilizing a neural network implemented in Dragonfly~\cite{dragonfly}. 
More precisely, a 2.5D U-net++~\cite{Zhou2018} of depth level 5, initial filter count of 64, and accounting for 3 slices was trained on several hand-labeled cutouts.
Note that the studied PNFs and PAs exhibit large differences in their morphologies, which increases the difficulty of training a single neural network that is capable of segmenting them all. Therefore, we decided to retrain individual neural networks based on one common pre-trained network for segmenting each tomogram.
Exemplary cross sections of the phase-wise segmentations for each peptide are shown in Figures~\ref{fig:peptidesOverview}(e)-(h).

Subsequently, the aggregate-wise segmentations are derived utilizing the \textit{label} function of scikit-image~\cite{VanderWalt2014}, which assigns each disconnected component a unique label. 
Mathematically, the aggregate-wise segmentation 
$\peptideWise: W\to\{0,1,\ldots,n\}$
can be expressed by
\begin{align*}
    \peptideWise(x)=
    \begin{cases}
        i, \quad &\text{if $x$ corresponds to the $i$-th aggregate,}\\
        0, &\text{else,}
    \end{cases}
\end{align*}
for each $x\in\W$.
Further, the set of aggregates is denoted by $\peptideSet=\{\peptide_1,\ldots,\peptide_{n}\}$, where  $\peptide_i=\{x\in\W\colon\peptideWise(x)=i\}$ for $i=1,\ldots,n$ and $n\in\N$ denotes the number of aggregates.
Analogously, the virion-wise segmentation $\virionWise$ and the set of virions $\virionSet$ are defined.

\begin{figure}[H]
    \centering
    \includegraphics[width=.9\textwidth]{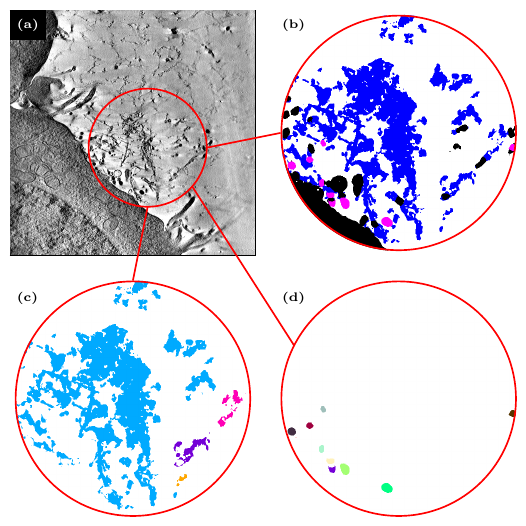}
    \caption{First, the STEM image $\STEM$ (a) is segmented phase-wise (b), into cell (black), peptide (blue) and virions (magenta).
    Subsequently, the phases are segmented instance-wise, i.e., each disconnected component of the peptide phase (c) and virion phase (d) is assigned a unique label, represented by different colors. 
    Note that disconnected components in the shown 2D cross sections with the same color are connected in 3D.
    }
    \label{fig:segmentation}
\end{figure}

Note that disconnected components of the cell and peptide phases with a volume smaller than 1000 voxels ($\approx$ 0.57\textmu m$^3$), as well as connected components of the virion phase smaller than 250 voxels ($\approx 1.5\cdot10^{-4}$\textmu m$^3$), are considered noise and assigned to the background phase.
These volume thresholds were chosen heuristically and were found to provide a good balance between noise reduction and the false removal of cell, peptide, and virion components.
Exemplary cross sections of the phase-wise, aggregate-wise, and virion-wise segmentations are presented in Figure~\ref{fig:segmentation}.

The application of the outlined procedure to all tomograms results in 17 segmented data sets corresponding to four peptides. 
Further details on the segmented tomograms are listed in Table~\ref{tab:SegmentationSpecifications}.

\begin{table}[H]
    \centering
    \begin{tabular}{cccccc}
    \toprule
    Peptide & \makecell{Number of\\[-6pt] cutouts} & \makecell{Number of\\[-6pt] aggregates} & \makecell{Number of\\[-6pt] virions} & \makecell{Mean cell \\[-6pt]volume fraction}\\
    \midrule
        D4 & 3 & 216 & 136 &  0.196\\
        eic-PA & 6 & 161 & 87 & 0.368\\
        pal-PA & 4 & 126 & 286 & 0.392\\
        Vect-1 & 4 & 126 & 78 & 0.341\\
        \bottomrule
    \end{tabular}
    \caption{Overview of the details of all segmented tomograms. The mean cell volume fraction indicates how much of the tomogram  is occupied by the cell phase on average. Large fractions imply less volume for peptides and virions.}
    \label{tab:SegmentationSpecifications}
\end{table}

\subsection{Geometric descriptors}\label{sec:Descriptors}
The previously determined phase- and instance-wise segmentations allow the computation of several geometric descriptors that reduce specific properties of the complex voxelized 3D morphologies to scalar-valued quantities.
First, descriptors are defined that describe the morphology of peptide aggregates. 
Subsequently, descriptors for the interactions between peptide aggregates, virions, and cell phases are introduced.
Note that the selection of the chosen descriptors is discussed in Section~\ref{ssec:edgeEffects}.
These descriptors allow the determination of empirical probability densities, which will be analyzed in Section~\ref{sec:Discussion}.
The probability densities are estimated utilizing the kernel density estimators as implemented by the python package \textit{seaborn}~\cite{Waskom2021}, which is based on \textit{SciPy}~\cite{Pauli2020}. 
The bandwidth  was chosen utilizing Scott's rule of thumb~\cite{scott2015}.

\paragraph{Specific surface area.}
The so-called specific surface area $\ssa(\peptide)$ of an aggregate $\peptide\in\peptideSet$ represents the surface area per unit-volume of $\peptide$ and is given by 
\begin{align*}
    \ssa(\peptide)=\frac{\surfArea(\peptide)}{\volume(\peptide)},
\end{align*}
where $\surfArea(\peptide)$ denotes the area and $\volume(\peptide)$ the volume of $\peptide$.
For the discretized image data considered here, the value of $\surfArea(\peptide)$ is estimated  by the algorithm presented in~\cite{ohser2009}, and $\volume(\peptide)$ is determined by the number of voxels corresponding to $\peptide$.

\begin{figure}[H]
    \centering
    \includegraphics[width=0.5\linewidth]{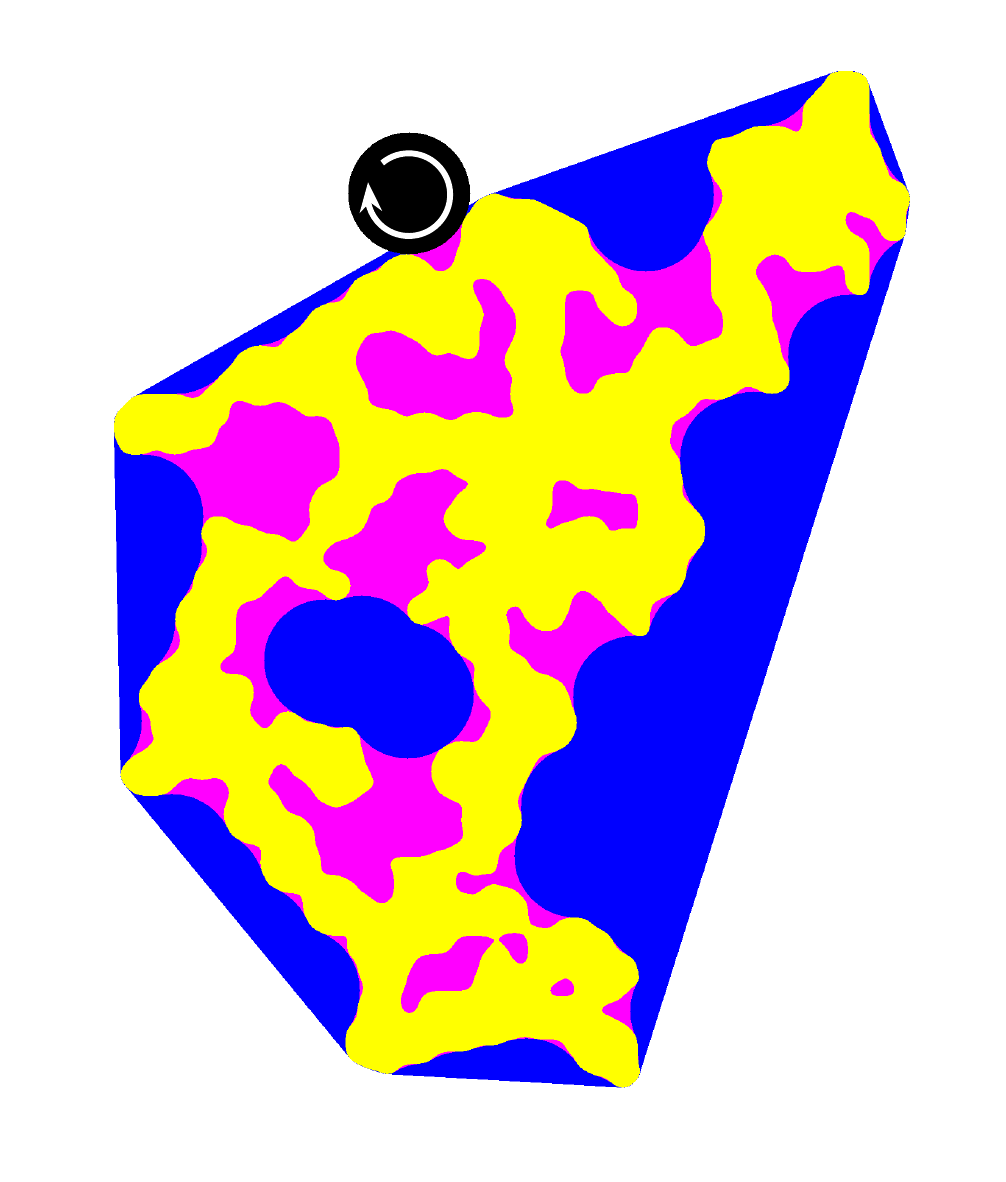}
    \caption{2D sketch of the rolling ball algorithm. 
    Yellow indicates an peptide aggregate, magenta represents the rolling-ball hull and blue the convex hull. 
    Black indicates an exemplary ball, which was used to derive the rolling-ball hull.
    }
    \label{fig:rollingBall}
\end{figure}

\paragraph{Volumetric occupancy.}
The volumetric occupancy of a general object is defined as its volume, normalized by the volume of a certain enclosing domain.
Frequently, the convex hull is chosen as such an enclosing domain.
In such cases, the volumetric occupancy is commonly referred to as \enquote{convexity} or \enquote{solidity}, see~\cite{RIEDER2025115602,FUCHS2026121475}. 
For the analysis of highly non-convex shapes, like the aggregates in the present application, the convex hull turned out to be too large, i.e., not tightly enclosing the aggregate.
This would lead to very low convexity values for all tomograms, which hinders interpretability.
A more reasonable choice for comparing the peptides considered in this paper is a so-called rolling-ball hull, derived by a rolling ball approach~\cite{sternberg1983Biomedical}.
Figuratively  speaking, a ball of radius $r>0$ rolls within the void space on the surface of an aggregate $\peptide\in\peptideSet$, see Figure~\ref{fig:rollingBall}. 
The rolling-ball hull is defined as the set of all voxels that cannot be covered by this ball. 
This algorithm is implemented by the morphological closing~\cite{soille1999} of $\peptide$ with a ball-shaped structuring element of radius $r$.
Mathematically expressed, the volumetric occupancy $\density(\peptide)$ of $\peptide$ is given as
\begin{align*}
    \density(\peptide)=\frac{\volume(\peptide)}{\volume\bigl(\text{close}(\peptide,r)\bigr)
    },
\end{align*}
where $\text{close}(\cdot\,,r)$ denotes the morphological closing with a ball-shaped structuring element of radius $r$. 
A value of $\density(\peptide)=1$ corresponds to an aggregate that matches its rolling-ball hull, while smaller values of $\density(\peptide)$ indicate less dense shapes.
It is important to note that, in contrast to a convex hull, the rolling-ball hull may have holes.\\
Since the diameter of a typical virion in the tomograms is between $\SI{100}{\nano\meter}$ and $\SI{150}{\nano\meter}$ (radius between $\SI{50}{\nano\meter}$ and $\SI{75}{\nano\meter}$), the rolling-ball radius $r$ was chosen as $r = 10$ voxels ($\widehat{=}\ \SI{83}{\nano\meter}$). This ensures that cavities within the aggregate that are accessible to virions are preserved.

\paragraph{Coverage ratio.}
Let $\peptide\in\peptideSet$ denote an aggregate. Then the coverage ratio $\coverageRatio(\peptide,\cellPhase)$ of $\peptide$ by the cell phase $\cellPhase$ is defined as
\begin{align*}
    \coverageRatio(\peptide,\cellPhase)=\frac{\surfArea(\peptide\cap\cellPhase)}{\surfArea(\peptide)},
\end{align*}
where $\surfArea(\peptide\cap\cellPhase)$ denotes the area of the interface between $\peptide$ and $\cellPhase$, and $\surfArea(\peptide)$ denotes the total surface area of $\peptide$.
A value of $\coverageRatio(\peptide,\cellPhase)=1$ implies that $\peptide$ is completely covered by the cell phase, whereas decreasing values correspond to smaller covered surface fractions. 
The joint surface $\coverageRatio(\peptide,\cellPhase)$ is numerically approximated by
\begin{align*}
    \coverageRatio(\peptide,\cellPhase)=\frac{\surfArea(\peptide)+\surfArea(\cellPhase)-\surfArea(\peptide\cup\cellPhase)}{2},
\end{align*}
utilizing the algorithm presented in~\cite{ohser2009}.
Analogously, the coverage ratio of a virion $\virion\in\virionSet$ by the cell phase $\cellPhase$ as well as the peptide phase $\peptidePhase$ is denoted as $\coverageRatio(\virion,\cellPhase)$ and $\coverageRatio(\virion,\peptidePhase)$, respectively.

\paragraph{Minimum distance.}
Let $\virion\in\virionSet$ denote a virion and $\cellPhase$ the cell phase. 
Then, the minimum distance $\minDist(\virion,\cellPhase)$ between $\virion$ and $\cellPhase$ is given as
\begin{align*}
    \minDist(\virion,\cellPhase)= \min_{x\in \virion,y\in\cellPhase}|x-y|,
\end{align*}
where $|\cdot|$ denotes the Euclidean norm.
For the practical implementation, let $D_\cellPhase\colon\W\rightarrow[0,\infty)$ denote the Euclidean distance transform from the cell phase to its complement, i.e., each voxel $x\in\W$ is assigned its minimum distance to $\cellPhase$~\cite{Russ2006}.
Then, the minimum distance $\minDist(\virion,\cellPhase)$ between $\virion$ and $\cellPhase$ is computed as
\begin{align}
    \minDist(\virion,\cellPhase)= \min\{D_\cellPhase(x)-1\colon x\in\virion\}.
    \label{eq:minDist}
\end{align}
Since $D_\cellPhase(x)\geq1$ for any voxel $x\notin\cellPhase$, the subtraction by $1$ in Eq.~\eqref{eq:minDist} ensures that a virion $\virion$ in direct contact with $\cellPhase$ yields a minimum distance of $\minDist(\virion,\cellPhase) = 0$.

Analogously, the minimum distance between a virion $v$ and the peptide phase $\peptidePhase$ is denoted by $\minDist(\virion,\peptidePhase)$ and defined similarly.

\section{Results and Discussion}\label{sec:Discussion}

\subsection{Quantitative description of peptide aggregate morphology}
To objectively compare peptide-virion-cell interactions across the four peptides, we quantify morphological, spatial, and interfacial descriptors from segmented STEM tomograms. These descriptors capture complementary aspects of fibril aggregate structure, virion binding, and cell association, all of which may impact transduction-enhancing activity. 

Initially, we focus on characterizing the morphology of the peptide aggregates themselves, quantifying the specific surface area and volumetric occupancy. 
The latter also provides an estimate for the accessibility of cavities in the aggregate structure to viral particles. 
While volumetric occupancy is primarily relevant for accessibility to viral particles, the specific surface area is crucial for both virion binding and subsequent interactions with the cell surface, as it determines how much peptide material is exposed to multivalent electrostatic and hydrophobic interactions.

\begin{figure}[H]
    \centering
    \includegraphics[width=\linewidth]{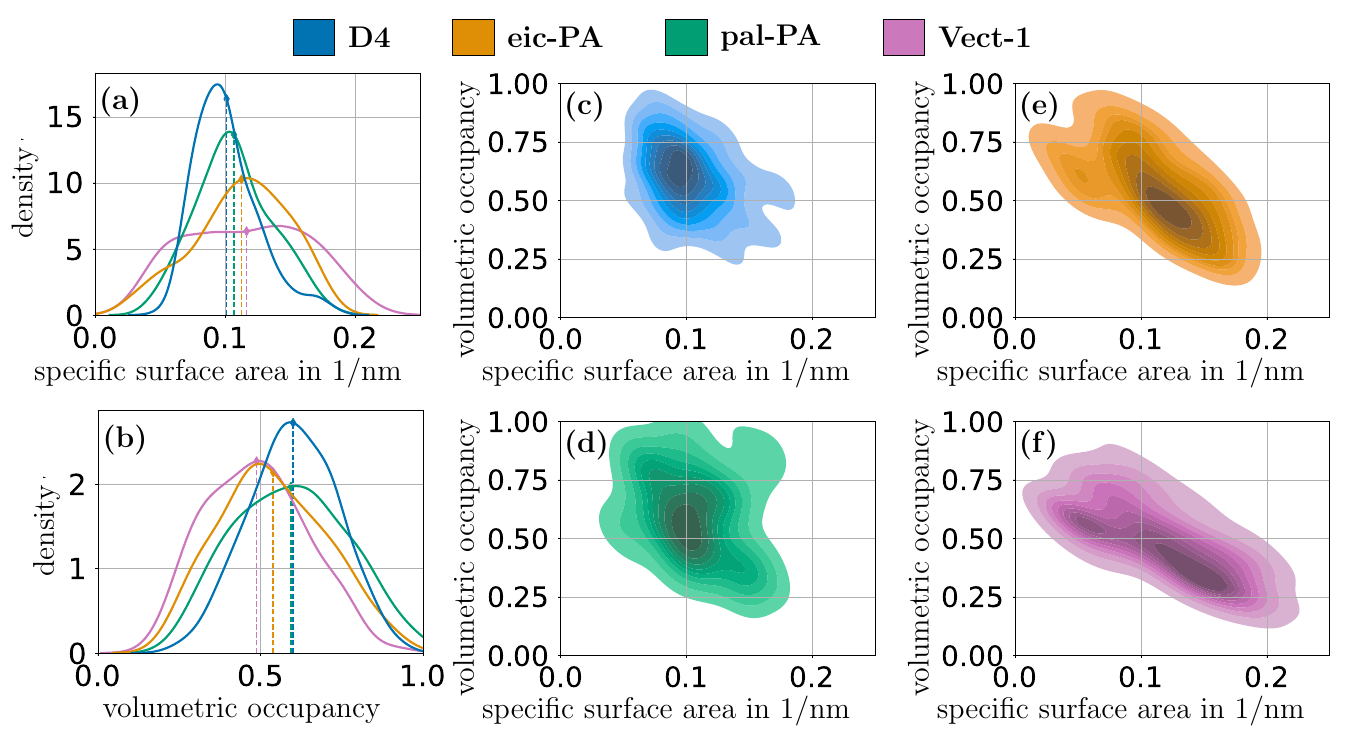}
    \caption{
    Univariate probability densities ((a) and (b)) of the specific surface area and the volumetric occupancy, along with the bivariate probability densities ((c)-(f)) of these two descriptors for each investigated peptide. The vertical lines indicate the mean values of the corresponding probability densities.  
    }
    \label{fig:peptideDescriptors}
\end{figure}

The specific surface area shows the highest values for Vect-1, followed by eic-PA, pal-PA, and lastly D4, see Figure~\ref{fig:peptideDescriptors}(a).
Vect-1 displays the lowest volumetric occupancy with relatively large heterogeneity, as indicated by the broader curve.
Conversely, D4 appears slightly denser but more homogeneously packed, see Figure~\ref{fig:peptideDescriptors}(b).
Pal-PA shows intermediate volumetric occupancy values, whereas eic-PA aggregates are characterized by lower density and a broader, more branched organization, reflecting loosely organized and heterogeneous assemblies. 
Figures~\ref{fig:peptideDescriptors}(c)-(f) show the bivariate probability densities of the specific surface area and the volumetric occupancy, implying a negative correlation between these two descriptors  for all four peptides.
However, the negative correlation is more pronounced for eic-PA and Vect-1 (Figures~\ref{fig:peptideDescriptors}(e) and (f)), suggesting that their assemblies undergo more cooperative transitions between open, high-surface-area morphologies and denser, more tightly packed structures.

Based on these descriptors, as well as most of the following ones, the peptides can be categorized into two groups with similar behavior: 
D4 and pal-PA exhibiting a relatively compact arrangement, while eic-PA and Vect-1 show a more open/porous morphology.

\subsection{Spatial organization of cell-peptide-virion interactions offer mechanistic insight into transduction enhancement}
\begin{figure}[H]
    \centering
    \includegraphics[width=\linewidth]{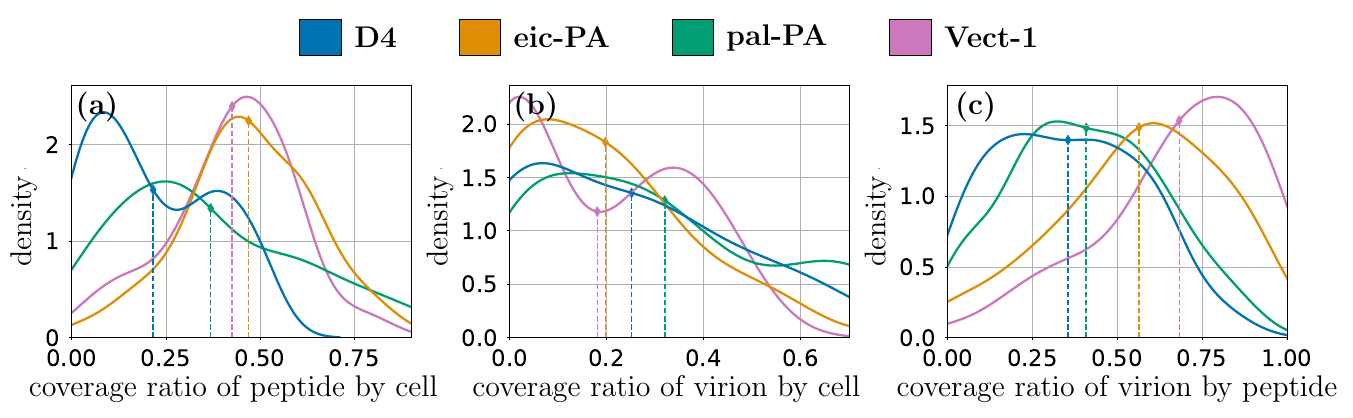}
    \caption{Coverage ratio of aggregates by cell, virions by cell and virions by peptide, conditioned on joint surfaces $>0$. The vertical lines indicate the mean values of the corresponding probability densities. 
    }
    \label{fig:univJointSurf}
\end{figure}

Interactions between peptide aggregates, cell surface, and virions are considered in the analysis of the coverage ratio of aggregates and virions, see Figure~\ref{fig:univJointSurf}.
Intuitively, dense D4 displays the lowest coverage ratio by the cell surface, while Vect-1 and eic-PA are substantially more covered by the cell surface than D4 and pal-PA (Figure~\ref{fig:univJointSurf}(a)). 
This is likely a result of the peptide morphologies (globular vs. more fibrillar). More globular peptides seem to be bound passively by filopodia, creating a mesh-like network that lies above the cells, while fibrillar structures are bound more actively by the cell, as membrane ruffling and engulfment suggest, see~\cite{LaRoche2026D4}.\\
While the same trend was observed for virions coverage by peptide (Figure~\ref{fig:univJointSurf}(c)), indicating that globular peptides exhibit enhanced binding to virions compared to more fibrillar peptides, this does not result in clear effects on the coverage ratio of virions by cells (Figure~\ref{fig:univJointSurf}(b)). 
However, the biological meaning of \enquote{coverage of virions by cell} as numeric values is limited, as it represents a static snapshot of a dynamic, multi-step entry process. 
Indeed, D4 is actively engulfed by cells, while Vect-1 binds to filopodia in a rather passive fashion~\cite{LaRoche2026D4}. Virion uptake in the presence of peptide enhancers is expected to occur predominantly via internalization of peptide-virion assemblies, which can be tethered to the plasma membrane at limited contact sites while most virions remain spatially separated from the membrane. Therefore, the more informative readouts are the extent of virion-peptide association (viral capture) and peptide-cell contact (cell engagement). 

\newpage

\begin{figure}[h!]
    \centering
    \includegraphics[width=\linewidth]{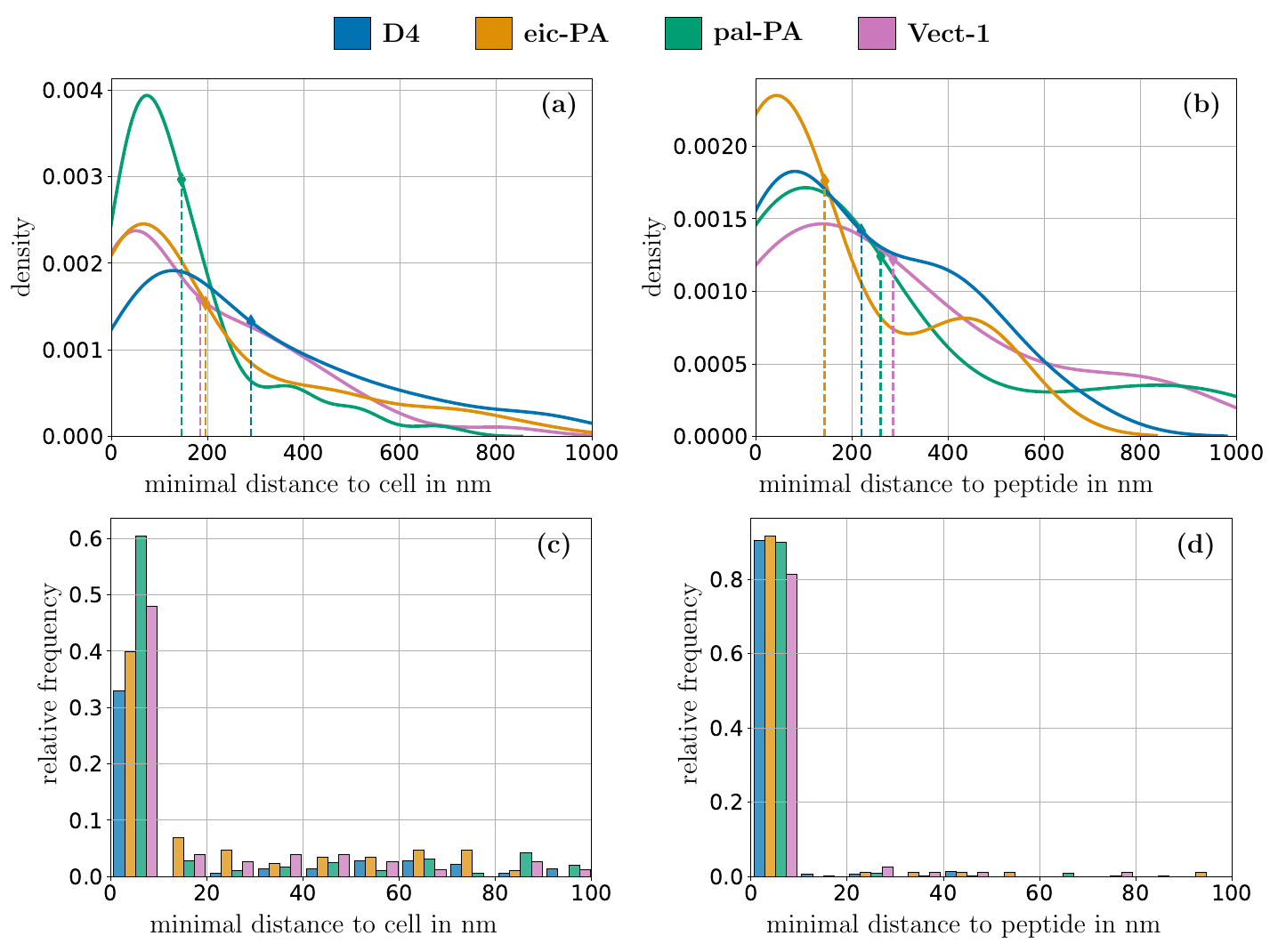}
    \caption{Probability density of the minimum distances of virions to cell phase (a) and peptide phase (b), conditioned on distance larger than zero.
    The vertical lines indicate the mean values of the corresponding probability densities. 
    Histograms showing the relative frequency of distances of virions to cell phase (c) and peptide phase (d), including distances equals to zero.
    Note that the y-axes are differently scaled.
    }
    \label{fig:minDists}
\end{figure}

Figures~\ref{fig:minDists}(a) and (b) show the minimum distance of virions to the cell and to the peptide, respectively.
Here, only virions not in direct contact with the peptide (distance $>$ 0) are considered, since virions at zero distance are already captured by the coverage ratio analysis above. Importantly, virions are rarely found at large distances from either the cell or peptide in any condition, reflecting efficient sequestration by all four peptides (within the narrow range of observation). 
Notably, the minimum distance to the cell (Figure~\ref{fig:minDists}(a)) does not follow the two-group pattern observed for aggregate morphology.
While aggregate architecture separates into compact (D4, pal-PA) and open/porous structures (eic-PA, Vect-1), virion-to-cell distances show a clear difference between D4 and pal-PA, whereas eic-PA and Vect-1 exhibit similar behavior.
For the minimum distance to peptide (Figure~\ref{fig:minDists}(b)), eic-PA displays the shortest distances, likely because eic-PA aggregates extensively coat the cell surface, placing peptide material in close proximity to cell-associated virions. Figures~\ref{fig:minDists}(c) and (d) include all virions irrespective of distance, confirming that the majority are located in the immediate vicinity of both cells and peptide aggregates.

\begin{figure}[h!]
    \centering
    \includegraphics[width=\linewidth]{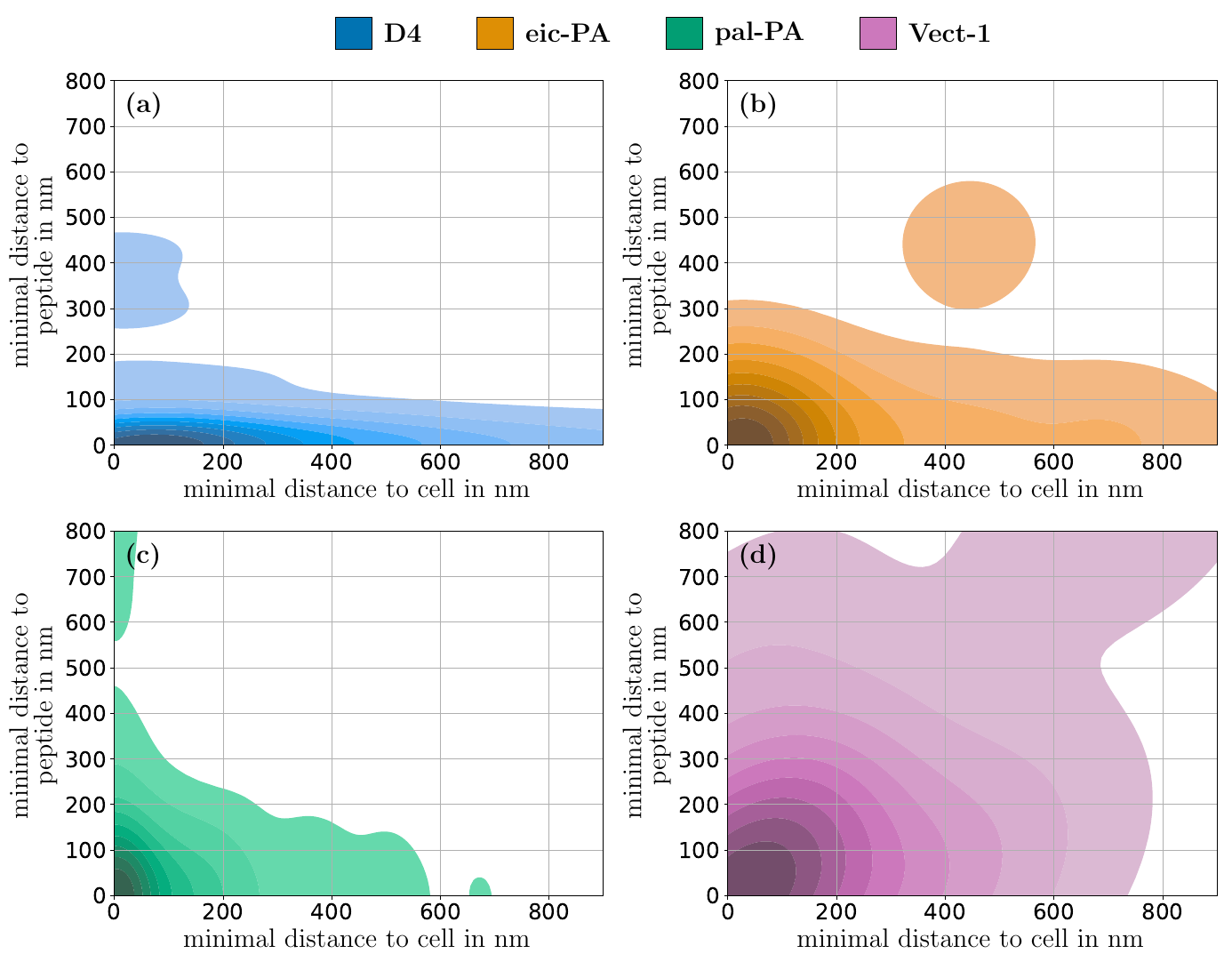}
    \caption{Bivariate probability density of minimum distances of a virion to cell and peptide.
    }
    \label{fig:minDistsBiv}
\end{figure}

Lastly, the bivariate probability densities of the minimum distance from virions to the cell and to the peptide are analyzed, see Figure~\ref{fig:minDistsBiv}.
For D4, pal-PA, and eic-PA, virions are almost exclusively confined to regions close to the cell, the peptide, or both, indicating effective spatial trapping within a peptide-cell interaction zone. 
In contrast, Vect-1 samples additionally contain a small population of virions located at larger distances from both the cell and peptide, suggesting less complete spatial confinement. 
These data indicate that while all peptides enhance viral capture, D4, pal-PA, and eic-PA more efficiently restrict virions to the near-cell environment, consistent with aggregate-mediated or proximity-driven uptake mechanisms.

Importantly, all peptides analyzed here are established transduction enhancers, and the spatial analysis does not contradict their functional efficacy.
Rather than rating enhancers by overall performance, the data reveal that distinct aggregate architectures result in transduction enhancement through different spatial organization. 
All peptides efficiently capture virions and result in few free virions. 
However, they differ in how strictly virions are confined to the near-cell environment. Denser or amorphous assemblies (D4, pal-PA, and eic-PA) promote tighter spatial confinement, while more porous Vect-1 displays a broader spatial distribution of virions, including a small fraction located at larger distances from both the cell and peptide. Importantly, all of the herein presented peptides are capable of substantially enhancing lentiviral transduction, albeit with varying potency.

\begin{figure}[H]
    \centering
    \includegraphics[width=0.6\linewidth]{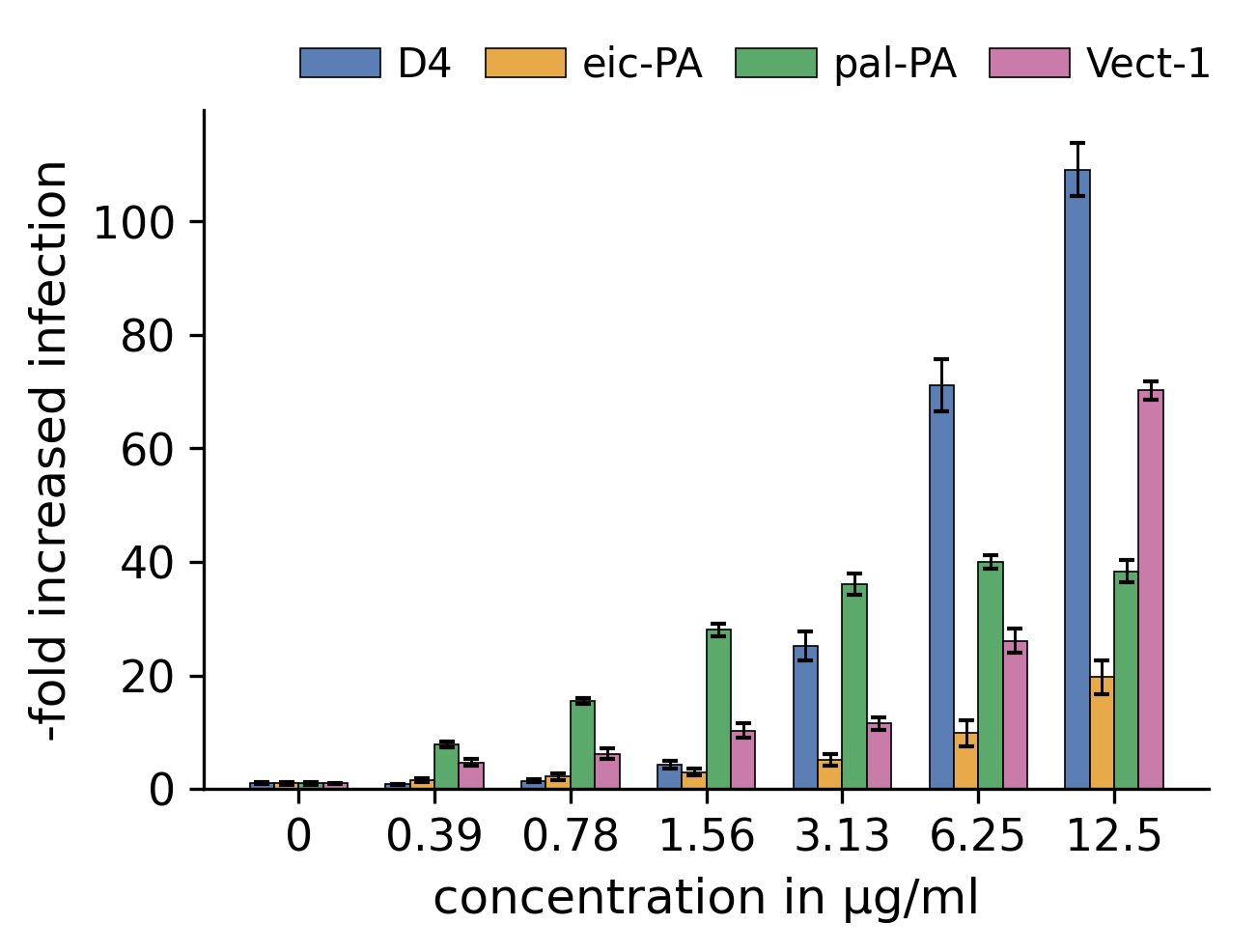}
    \caption{   
    Enhancement of HIV-1 infection (NL4-3 R5) by peptide nanofibrils (PNFs) and peptide amphiphiles (PAs). Peptides were incubated with virions for 10 min at room temperature at the indicated concentrations prior to addition to TZM-bl cells. Infection levels were quantified after 48 h using a $\beta$-galactosidase reporter assay and are reported as relative light units per second (RLU/s).
    The height of the histogram represents the mean values of three independent experiments performed in triplicates, while the error bars visualize the standard deviation.
    }
    \label{fig:EnHancement}
\end{figure}

As demonstrated for HIV-1 infection (Figure~\ref{fig:EnHancement}), D4 achieved the highest relative enhancement of infection after treating virus stocks with  \SI{12.5}{\micro\gram/\milli\liter} of the peptide (122-fold), followed by Vect-1 (76-fold), pal-PA (30-fold), and eic-PA (20-fold). It is conceivable that excessively tight virion confinement (as seen for pal-PA and eic-PA) could become counterproductive by limiting access to the cell membrane or restricting entry-related dynamics. 
Although this cannot be resolved from the present static analysis, our data are consistent with transduction enhancement arising from a balance between strong virion capture and sufficient peptide-virion assembly flexibility. 
Beyond the specific peptides analyzed here, this work establishes a mathematical framework, based on 3D STEM tomograms, for quantitatively evaluating peptide-virus-cell interactions. By linking aggregate morphology, spatial organization, and interfacial contacts, this approach provides a generalizable tool to guide the rational design of next-generation peptide nanofibrils and amphiphiles for transduction enhancement.

\subsection{Edge effects, sampling bias, and descriptor selection}\label{ssec:edgeEffects}
A well-known issue in computer vision arises when image data do not capture the entire region of interest: descriptors derived from such data can be biased by edge effects, which occur when objects are truncated by the observation window.
A common strategy to mitigate this bias is the so-called minus-sampling, where only objects that are entirely contained within the observation window are included in subsequent analyses~\cite{chiu2013}.

However, since the probability that an object is intersected by the observation window is proportional to its size~\cite{baddeley2004}, small peptide aggregates are more likely to be fully contained than large ones.
While this effect may be negligible for large observation windows, in the present case, it would drastically reduce the number of available aggregates—by up to $\approx70\%$ (see Table~\ref{tab:comparison}), due to the thin observation windows ($\approx\SI{0.5}{}-\SI{1}{\micro\meter}$ in thickness). 
Increasing the thickness of the section and consequently increasing the database is not feasible, as thicker sections in STEM tomography intensify electron–matter interactions, causing beam broadening and multiple scattering, which degrade contrast and spatial resolution~\cite{hohn_preparation_2011, aoyama_stem_2008}.

As a consequence of excluding intersected aggregates, the probability mass of the aggregate size distribution would be erroneously shifted toward smaller aggregates. 
This effect is illustrated in Figures~\ref{fig:comparison}(a) and (b),
where the volume-equivalent diameter of peptide aggregates and that of their rolling-ball hulls (both representing aggregate size) are compared for all aggregates and for those entirely contained within the observation window.
Due to potential (hidden) correlations, this so-called sampling bias may also affect other descriptors.

In view of these issues, only descriptors that are normalized by the size of individual aggregates are considered in the present work.
In general, such normalized descriptors may still be biased, as they could be correlated with aggregate size.
However, in the present application, the specific surface area and the volumetric occupancy exhibit similar probability densities when comparing all peptides with those that are entirely contained in the observation window, see Figures~\ref{fig:comparison}(c) and (d).
This observation indicates that these normalized descriptors are largely independent of aggregate size, motivating the assumption that these probability distributions obtained from all aggregates are representative and less affected by edge effects.

\section{Conclusion}
In this study, we introduced a quantitative framework for analyzing peptide-virion-cell interactions based on segmented STEM tomography datasets and geometric descriptors. Comparing four infection-enhancing peptides, we show that all efficiently capture virions but differ in aggregate geometry and how strictly they confine particles near the cell surface. Compact, dense aggregates (D4, pal-PA) promote tight virion confinement at the peptide-cell interface, correlating with stronger transduction enhancement, whereas porous (Vect-1) or amorphous (eic-PA) architectures distribute virions more broadly. These findings suggest that compact confinement near the membrane is a key biophysical determinant of enhancer efficacy, although biological aspects of cellular uptake likely extend beyond the parameters investigated here. A limitation is the relatively thin observation window inherent to STEM tomography sections, which introduces edge effects for larger aggregates. However, STEM tomography provides the thickest sections achievable among electron tomography techniques while maintaining the spatial resolution required to resolve individual virions and peptide aggregate morphology. Volume imaging approaches such as Focused ion beam (FIB)-Scanning electron microscopy (SEM) or Serial block face (SBF)-SEM offer larger fields of view and volumes but lack the resolution needed for this type of analysis. 

Training a generalized segmentation model across peptide conditions could streamline future analyzes but would require substantially larger segmented datasets and is beyond the scope of the present study. 
Despite these limitations, the present dataset of 17 tomograms already provides new insights into the spatial organization of four structurally diverse peptides, supporting a link between aggregate architecture and transduction efficacy. This framework thus offers a quantitative basis to guide the rational design of next-generation transduction enhancers for therapeutic applications such as CAR-T cell manufacturing.

\section*{Acknowledgments}
We thank Renate Kunz for technical support. 
This project has received funding through the German Research Foundation (432000323) and through a Collaborative Research Center grant of the German Research Foundation (316249678 – SFB 1279) to CR and JM. 

\section*{Data Availability Statement}
The datasets generated and/or analyzed during the current study are available from the corresponding authors on reasonable request.

\bibliographystyle{myAbbrvnat}
\bibliography{references}

@article{aoyama_stem_2008,
	title = {{STEM} tomography for thick biological specimens},
	volume = {109},
	journal = {Ultramicroscopy},
	author = {Aoyama, Kazuhiro and Takagi, Tomoko and Hirase, Ai and Miyazawa, Atsuo},
	year = {2008},
	pages = {70--80},
}

@article{hohn_preparation_2011,
	title = {Preparation of cryofixed cells for improved {3D} ultrastructure with scanning transmission electron tomography},
	volume = {135},
	journal = {Histochemistry and Cell Biology},
	author = {Höhn, Katharina and Sailer, Michaela and Wang, Li and Lorenz, Myriam and Schneider, Marion and Walther, Paul},
	year = {2011},
	pages = {1--9},
}

@article{walther_freeze_2002,
	title = {Freeze substitution of high-pressure frozen samples: {T}he visibility of biological membranes is improved when the substitution medium contains water},
	volume = {208},
	urldate = {2023-06-19},
	journal = {Journal of Microscopy},
	author = {Walther, P. and Ziegler, A.},
	year = {2002},
	pages = {3--10},
}

@software{dragonfly,
    title = {Dragonfly},
    author = {{Comet Technologies Canada Inc.}},
    year = {2025},
    address = {Montreal, Canada},
note={{V}ersion: 2025.1}
}

@Article{VanderWalt2014,
  author    = {Van der Walt, Stefan and Sch{\"o}nberger, Johannes L and Nunez-Iglesias, Juan and Boulogne, Fran{\c{c}}ois and Warner, Joshua D and Yager, Neil and Gouillart, Emmanuelle and Yu, Tony},
  journal   = {PeerJ},
  title     = {scikit-image: {I}mage processing in {Python}},
  year      = {2014},
  pages     = {e453},
  volume    = {2},
  publisher = {PeerJ Inc.},
}

@Article{Patel2025,
  author  = {Patel, Kiran K. and Tariveranmoshabad, Mahsa and Kadu, Suyog and Shobaki, Nour and June, Carl H.},
  title   = {From concept to cure: {T}he evolution of {CAR}-{T} cell therapy},
  journal = {Molecular Therapy},
  year    = {2025},
  volume  = {33},
  pages   = {2123--2140},
}

@Article{RauchWirth2025,
  author  = {Rauch-Wirth, Lena and Sch{\"u}tz, Daniel and Gro{\ss}, R{\"u}diger and Rode, Sophie and Glocker, Benjamin and M{\"u}ller, Julian A. and Walther, Peter and Read, Clarissa and M{\"u}nch, Jan},
  title   = {Transduction enhancing {EF-C} peptide nanofibrils are endocytosed by macropinocytosis and subsequently degraded},
  journal = {Biomaterials},
  year    = {2025},
  volume  = {317},
  pages   = {123044},
}

@article{LaRoche2026D4,
title = {Structural morphology of peptide nanofibrils dictates viral capture and cellular uptake in gene therapy applications},
journal = {Nano Today},
volume = {69},
pages = {103015},
year = {2026},
author = {Rauch-Wirth, Lena and La Roche, Julia and Kaygisiz, Kübra and Kuhn, Annalena and Jeon, Nayeong and Stifel, Ulrich and Zech, Fabian and Fischer-Posovszky, Pamela and Weil, Tanja and Read, Clarissa and Münch, Jan},
}

@Article{Muench2007,
  author  = {M{\"u}nch, Jan and R{\"u}cker, Eva and St{\"a}ndker, Ludger and Adermann, Knut and Goffinet, C{\'e}cile and Schindler, Matthias and Wildum, Sophie and Chinnadurai, Ramesh and Rajan, Divya and Specht, Andreas and Gim{\'e}nez-Gallego, Guillermo and S{\'a}nchez, Pedro C. and Fowler, Douglas M. and Koulov, Alexei and Kelly, Jeffery W. and Mothes, Walther and Grivel, Jean-Charles and Margolis, Leonid and Keppler, Oliver T. and Forssmann, Wolf-Georg and Kirchhoff, Frank},
  title   = {Semen-derived amyloid fibrils drastically enhance {HIV} infection},
  journal = {Cell},
  year    = {2007},
  volume  = {131},
  pages   = {1059--1071}
}

@Article{Roan2009,
  author  = {Roan, Nadine R. and M{\"u}nch, Jan and Arhel, Nathalie and Mothes, Walther and Neidleman, Jason and Kobayashi, Akiko and Smith-McCune, Karen and Kirchhoff, Frank and Greene, Warner C.},
  title   = {The cationic properties of {SEVI} underlie its ability to enhance human immunodeficiency virus infection},
  journal = {Journal of Virology},
  year    = {2009},
  volume  = {83},
  pages   = {73--80},
}

@Article{Roan2014,
  author  = {Roan, Nadine R. and Liu, Hong and Usmani, Syed M. and Neidleman, Jason and M{\"u}ller, Julian A. and Avila-Herrera, Alexander and Gawanbacht, Armin and Zirafi, Omid and Chu, Steven and Dong, Min and Kumar, S. T. and Smith, Jason F. and Pollard, Katherine S. and F{\"a}ndrich, Marcus and Kirchhoff, Frank and M{\"u}nch, Jan and Witkowska, Hanna E. and Greene, Warner C.},
  title   = {Liquefaction of semen generates and later degrades a conserved semenogelin peptide that enhances {HIV} infection},
  journal = {Journal of Virology},
  year    = {2014},
  volume  = {88},
  pages   = {7221--7234},
}

@InProceedings{Zhou2018,
author="Zhou, Zongwei and Rahman Siddiquee, Md Mahfuzur and Tajbakhsh, Nima and Liang, Jianming", 
editor="Stoyanov, Danail and Taylor, Zeike and Carneiro, Gustavo
and Syeda-Mahmood, Tanveer and Martel, Anne and Maier-Hein, Lena
and Tavares, Jo{\~a}o Manuel R.S. and Bradley, Andrew and Papa, Jo{\~a}o Paulo and Belagiannis, Vasileios and Nascimento, Jacinto C. and Lu, Zhi and Conjeti, Sailesh and Moradi, Mehdi and Greenspan, Hayit and Madabhushi, Anant",
title="UNet++: {A} Nested {U}-Net Architecture for Medical Image Segmentation",
booktitle="Deep Learning in Medical Image Analysis and Multimodal Learning for Clinical Decision Support",
year="2018",
publisher="Springer",
pages="3--11",
}

@book{ohser2009,
  title={3{D} {I}mages of {M}aterials {S}tructures},
  author={Ohser, Joachim and Schladitz, Katja},
  year={2009},
  publisher={Wiley-VCH}
}

@ARTICLE{sternberg1983Biomedical,
  author={Sternberg, S. R.},
  journal={Computer}, 
  title={Biomedical Image Processing}, 
  year={1983},
  volume={16},
  pages={22-34},
}

@book{soille1999,
  title={Morphological Image Analysis: {P}rinciples and Applications},
  author={Soille, Pierre},
  year={1999},
  publisher={Springer}
}

@article{RIEDER2025115602,
title = {Statistical analysis of grains and pores within polycrystalline {Al$_2$TiO$_5$} ceramics, based on {X}-ray computed tomography},
journal = {Materials Characterization},
volume = {229},
pages = {115602},
year = {2025},
author = {Philipp Rieder and Lukas Petrich and Itziar Serrano-Munoz and Mossaab Mouiya and Henning Markötter and Marc Huger and Giovanni Bruno and Volker Schmidt},
}

@article{LaRoche2025,
author = {La Roche, Julia and Rauch-Wirth, Lena and Zimmerman, Laura and Zech, Fabian and Münch, Jan and Read, Clarissa and Kaygisiz, Kübra},
title = {Interactions of Peptide Amphiphiles With Viruses and Cells Are Enabled by Amorphous Nanostructures},
journal = {Journal of Peptide Science},
volume = {31},
pages = {e70051},
year = {2025}
}

@article{Kaygisiz2024,
author = {Kaygisiz, Kübra and Rauch-Wirth, Lena and Iscen, Aysenur and Hartenfels, Jan and Kremer, Kurt and Münch, Jan and Synatschke, Christopher V. and Weil, Tanja},
title = {Peptide Amphiphiles as Biodegradable Adjuvants for Efficient Retroviral Gene Delivery},
journal = {Advanced Healthcare Materials},
volume = {13},
pages = {2301364},
year = {2024}
}

@article{Pashuck2010,
author = {Pashuck, E. Thomas and Cui, Honggang and Stupp, Samuel I.},
title = {Tuning Supramolecular Rigidity of Peptide Fibers through Molecular Structure},
journal = {Journal of the American Chemical Society},
volume = {132},
pages = {6041-6046},
year = {2010},
}

@article{Paramonov2006,
author = {Paramonov, Sergey E. and Jun, Ho-Wook and Hartgerink, Jeffrey D.},
title = {Self-Assembly of Peptide-Amphiphile Nanofibers: {T}he Roles of Hydrogen Bonding and Amphiphilic Packing},
journal = {Journal of the American Chemical Society},
volume = {128},
pages = {7291-7298},
year = {2006},
}

@article{wieland_scanning_2025,
	title = {Scanning Transmission Electron Microscopy Tomography in Virology: {3D} Imaging of High-pressure Frozen, Freeze-substituted Samples},
	journal = {Journal of Visualized Experiments},
	author = {Wieland, Johannes Georg and La Roche, Julia and Bergner, Tim and Habisch, Rebecca and Puschmann, Eva and Soza-Ried, Jorge and Schütz, Desiree and Münch, Jan and Walther, Paul and Dass, Martin and Read, Clarissa},
	year = {2025},
	pages = {68568},
    volume={222}
}

@article{radek_vectofusin-1_2019,
	title = {Vectofusin-1 Improves Transduction of Primary Human Cells with Diverse Retroviral and Lentiviral Pseudotypes, Enabling Robust, Automated Closed-System Manufacturing},
	volume = {30},
	journal = {Human Gene Therapy},
	author = {Radek, Constanze and Bernadin, Ornellie and Drechsel, Katharina and Cordes, Nicole and Pfeifer, Rita and Sträßer, Pia and Mormin, Mirella and Gutierrez-Guerrero, Alejandra and Cosset, François-Loïc and Kaiser, Andrew D. and Schaser, Thomas and Galy, Anne and Verhoeyen, Els and Johnston, Ian C. D.},
	year = {2019},
	pages = {1477--1493},
}

@article{rauch-wirth_optimized_2023,
	title = {Optimized peptide nanofibrils as efficient transduction enhancers for in vitro and ex vivo gene transfer},
	volume = {14},
	journal = {Frontiers in Immunology},
	author = {Rauch-Wirth, Lena and Renner, Alexander and Kaygisiz, Kübra and Weil, Tatjana and Zimmermann, Laura and Rodriguez-Alfonso, Armando A. and Schütz, Desiree and Wiese, Sebastian and Ständker, Ludger and Weil, Tanja and Schmiedel, Dominik and Münch, Jan},
	year = {2023},
	pages = {1270243},
}

@article{vermeer_vectofusin-1_2017,
	title = {\mbox{Vectofusin-1}, a potent peptidic enhancer of viral gene transfer forms {pH}-dependent $\alpha$-helical nanofibrils, concentrating viral particles},
	volume = {64},
	journal = {Acta Biomaterialia},
	author = {Vermeer, Louic S. and Hamon, Loic and Schirer, Alicia and Schoup, Michel and Cosette, Jérémie and Majdoul, Saliha and Pastré, David and Stockholm, Daniel and Holic, Nathalie and Hellwig, Petra and Galy, Anne and Fenard, David and Bechinger, Burkhard},
	year = {2017},
	pages = {259--268},
}

@article{irving_choosing_2021,
	title = {Choosing the Right Tool for Genetic Engineering: {C}linical Lessons from Chimeric Antigen Receptor-{T} Cells},
	volume = {32},
	journal = {Human Gene Therapy},
	author = {Irving, Melita and Lanitis, Evripidis and Migliorini, Denis and Ivics, Zoltán and Guedan, Sonia},
	year = {2021},
	pages = {1044--1058},
}

@article{hartgerink_self-assembly_2001,
	title = {Self-{assembly} and {mineralization} of {peptide}-{amphiphile} {nanofibers}},
	volume = {294},
	journal = {Science},
	author = {Hartgerink, Jeffrey D. and Beniash, Elia and Stupp, Samuel I.},
	year = {2001},
	pages = {1684--1688},
}

@article{cui_selfassembly_2010,
	title = {Self‐assembly of peptide amphiphiles: {From} molecules to nanostructures to biomaterials},
	volume = {94},
	journal = {Peptide Science},
	author = {Cui, Honggang and Webber, Matthew J. and Stupp, Samuel I.},
	year = {2010},
	pages = {1-18},
}

@book{baddeley2004,
  title={Stereology for {S}tatisticians},
  author={Baddeley, A. and Jensen, E.B.V.},
  year={2004},
  publisher={Taylor \& Francis}
}

@Book{chiu2013,
  author    = {S. Chiu and D. Stoyan and W. S. Kendall and J. Mecke},
  publisher = {J. Wiley \& Sons},
  title     = {Stochastic Geometry and its Applications},
  year      = {2013},
  edition   = {3rd},
}

@article{FUCHS2026121475,
title = {Stochastic modeling of particle structures in spray fluidized bed agglomeration using methods from machine learning},
journal = {Powder Technology},
volume = {467},
pages = {121475},
year = {2026},
author = {Lukas Fuchs and Sabrina Weber and Jialin Men and Niklas Eiermann and Orkun Furat and Andreas Bück and Volker Schmidt},
}

@article{weber2021multidimensional,
author = {Weber, Matthias and Wilhelm, Thomas and Schmidt, Volker},
year = {2021},
pages = {85-94},
title = {Multidimensional Characterisation of Time-dependent Image Data: {A} Case Study for the Peripheral Nervous System in Ageing Mice},
volume = {40},
journal = {Image Analysis and Stereology},
}

@article{BADIAMARTINEZ2012Three,
title = {Three-dimensional visualization of forming Hepatitis {C} virus-like particles by electron-tomography},
journal = {Virology},
volume = {430},
pages = {120-126},
year = {2012},
author = {Daniel Badia-Martinez and Bibiana Peralta and German Andrés and Milagros Guerra and David Gil-Carton and Nicola G.A. Abrescia},
}

@InProceedings{Ronneberger2015UNEt,
author="Ronneberger, Olaf
and Fischer, Philipp
and Brox, Thomas",
editor="Navab, Nassir
and Hornegger, Joachim
and Wells, William M.
and Frangi, Alejandro F.",
title="{U}-Net: {C}onvolutional Networks for Biomedical Image Segmentation",
booktitle="Medical Image Computing and Computer-Assisted Intervention -- MICCAI 2015",
year="2015",
publisher="Springer International Publishing",
pages="234--241",
}

@Article{Weber2023segmentation,
  author   = {Weber, Matthias and Neumann, Matthias and Schmidt, Matthias and Pfeiffer, Peter Benedikt and Bansal, Akanksha and Fändrich, Marcus and Schmidt, Volker},
  journal  = {Journal of Mathematics in Industry},
  title    = {Segmentation and morphological analysis of amyloid fibrils from cryo-{EM} image data},
  year     = {2023},
  pages    = {2},
  volume   = {13},
}

@article{heebner2022deep,
  title = {Deep Learning-Based Segmentation of Cryo-Electron Tomograms},
  author = {Heebner, Jessica E. and Purnell, Carson and Hylton, Ryan K. and Marsh, Mike and Grillo, Michael A. and Swulius, Matthew T.},
  year = {2022},
  volume = {189},
  pages = {e64435},
  journal = {Journal of Visualized Experiments},
}

@Article{McCraw2018structural,
  author   = {McCraw, Dustin M. and Gallagher, John R. and Torian, Udana and Myers, Mallory L. and Conlon, Michael T. and Gulati, Neetu M. and Harris, Audray K.},
  journal  = {Scientific Reports},
  title    = {Structural analysis of influenza vaccine virus-like particles reveals a multicomponent organization},
  year     = {2018},
  pages    = {10342},
  volume   = {8},
}

@article{martin2016distinct,
author = {Jessica L. Martin and Sheng Cao and Jose O. Maldonado and Wei Zhang and Louis M. Mansky},
title = {Distinct Particle Morphologies Revealed through Comparative Parallel Analyses of Retrovirus-Like Particles},
journal = {Journal of Virology},
volume = {90},
pages = {8074-8084},
year = {2016},
}

@Article{Maldonado2016DistinctMorphology,
AUTHOR = {Maldonado, José O. and Cao, Sheng and Zhang, Wei and Mansky, Louis M.},
TITLE = {Distinct Morphology of Human {T}-Cell Leukemia Virus Type 1-Like Particles},
JOURNAL = {Viruses},
VOLUME = {8},
YEAR = {2016},
pages = {132},
}

@book{Russ2006,
    author = {Russ, J.C.},
    title = {The Image Processing Handbook},
    publisher = {CRC Press},
    year = {2006},
    edition = {5th},
}

@article{Waskom2021,
    year = {2021},
    publisher = {The Open Journal},
    volume = {6},
    pages = {3021},
    author = {Michael L. Waskom},
    title = {Seaborn: {S}tatistical data visualization},
    journal = {Journal of Open Source Software}
 }

@ARTICLE{Pauli2020,
  author  = {Virtanen, Pauli and Gommers, Ralf and Oliphant, Travis E. and
            Haberland, Matt and Reddy, Tyler and Cournapeau, David and
            Burovski, Evgeni and Peterson, Pearu and Weckesser, Warren and
            Bright, Jonathan and {van der Walt}, St{\'e}fan J. and
            Brett, Matthew and Wilson, Joshua and Millman, K. Jarrod and
            Mayorov, Nikolay and Nelson, Andrew R. J. and Jones, Eric and
            Kern, Robert and Larson, Eric and Carey, C J and
            Polat, {\.I}lhan and Feng, Yu and Moore, Eric W. and
            {VanderPlas}, Jake and Laxalde, Denis and Perktold, Josef and
            Cimrman, Robert and Henriksen, Ian and Quintero, E. A. and
            Harris, Charles R. and Archibald, Anne M. and
            Ribeiro, Ant{\^o}nio H. and Pedregosa, Fabian and
            {van Mulbregt}, Paul and {SciPy 1.0 Contributors}},
  title   = {{{SciPy} 1.0: fundamental algorithms for scientific
            computing in python}},
  journal = {Nature Methods},
  year    = {2020},
  volume  = {17},
  pages   = {261--272}
}

@book{scott2015,
    title={Multivariate Density Estimation: {T}heory, Practice, and Visualization},
    author={Scott, David W},
    year={2015},
    publisher={J. Wiley \& Sons},
    edition={2nd}
}

\clearpage
\setcounter{figure}{0}
\setcounter{equation}{0}
\setcounter{table}{0}
\renewcommand{\theequation}{S\arabic{equation}}
\renewcommand{\thetable}{S\arabic{table}}
\renewcommand{\thefigure}{S\arabic{figure}}
\renewcommand*{\thesection}{S}

\section*{Supplementary Information} 
In Section~\ref{ssec:edgeEffects} edge effects are discussed, induced by the size of the observation window. 
Table~\ref{tab:comparison} shows the total number of peptide aggregates and virions, as well as the number of those intersected by the observation window.

\begin{table}[H]
    \centering
    \begin{tabular}{c| cc |cc| cc }
    \toprule
    &\multicolumn{2}{|c|}{Total} & \multicolumn{2}{|c|}{Intersected} & \multicolumn{2}{|c}{Fraction of intersected} \\
    \midrule
      & \makecell{number of\\[-6pt] peptide \\[-6pt]components} & \makecell{number of\\[-6pt] virions} & \makecell{number of\\[-6pt] peptide \\[-6pt]components} & \makecell{number of\\[-6pt] virions} & 
    peptides & virions \\
    \midrule
        D4      & 216   & 136   & 125    & 83    & 0.579     & 0.610\\
        eic-PA  & 161   & 87    & 88    & 49    &0.547      & 0.563\\
        pal-PA  & 156   & 286   & 74    & 146   &0.474      & 0.510\\
        Vect-1  & 126   & 78    & 89    & 45    & 0.706     & 0.577\\
        \bottomrule
    \end{tabular}
    \caption{Comparison of peptide aggregates and virions affected by edge effects.}
    \label{tab:comparison}
\end{table}

\begin{figure}[h!]
    \centering
    \includegraphics[width=\linewidth]{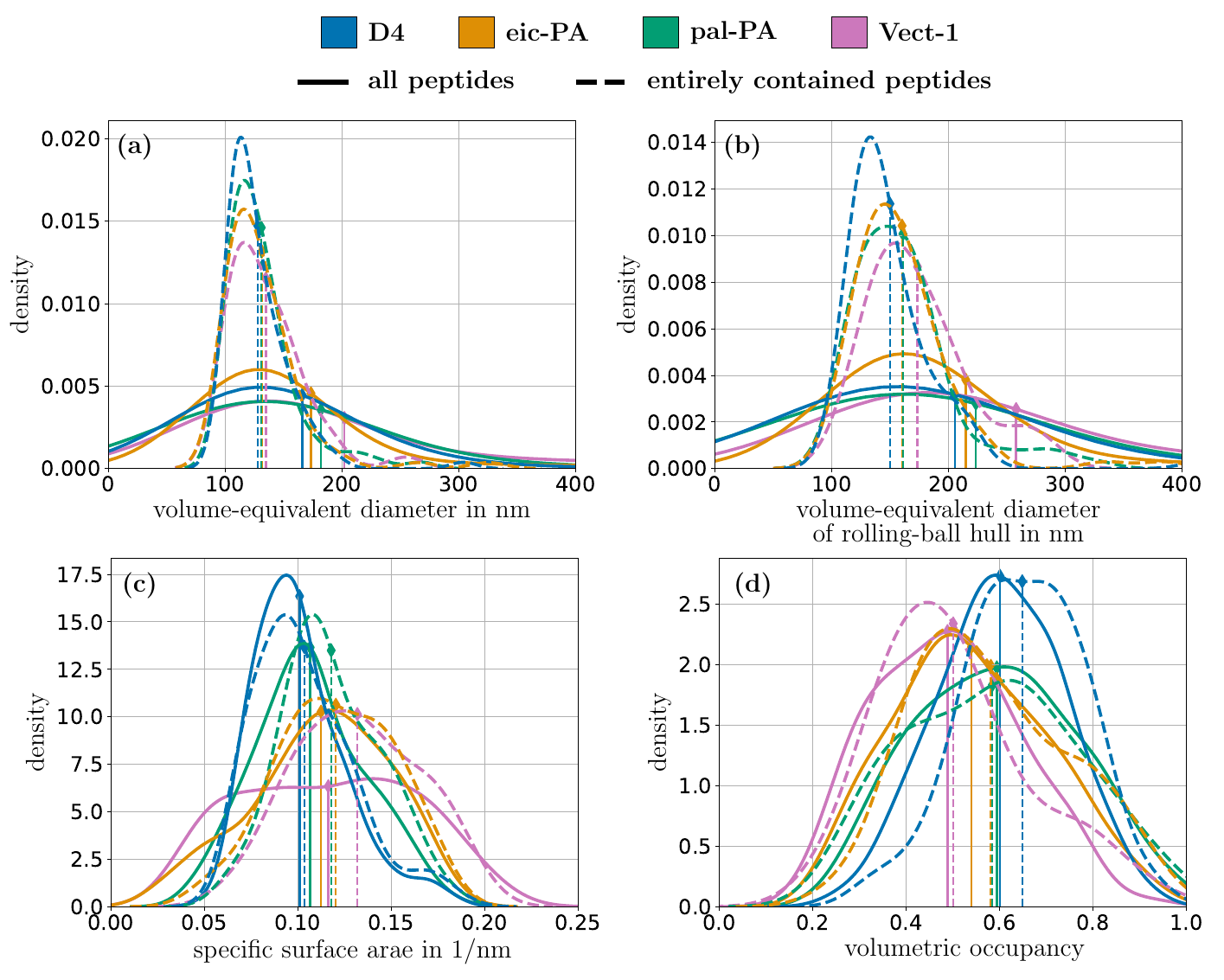}
    \caption{Comparison of probability densities of all peptides and those that are completely contained within the observation window $\W$.}
    \label{fig:comparison}
\end{figure}

To quantify edge effects onto the size distribution of peptide aggregates, let 
the size of a peptide $\peptide\in\peptideSet$ be represented by its volume-equivalent diameter $\ved(\peptide)$, which is formally defined as
\begin{align*}
    \ved(\peptide)=\sqrt[3]{\frac{6\volume(\peptide)}{\pi}}.
\end{align*}
Figures~\ref{fig:comparison}(a) and (b) show that the volume-equivalent diameter of aggregates and those of their rolling-ball hulls for all peptides, as well as those not intersected by the observation window, differ drastically.
However, normalized descriptors, such as the specific surface area and the rolling ball density (see Section~\ref{sec:Descriptors}), are less affected by such edge effects, since the curves corresponding to a peptide are similar.
Because of this, we assume that these descriptors are largely independent of aggregate size and are representative of all aggregates.

\begin{figure}
    \centering
    \includegraphics[width=\linewidth]{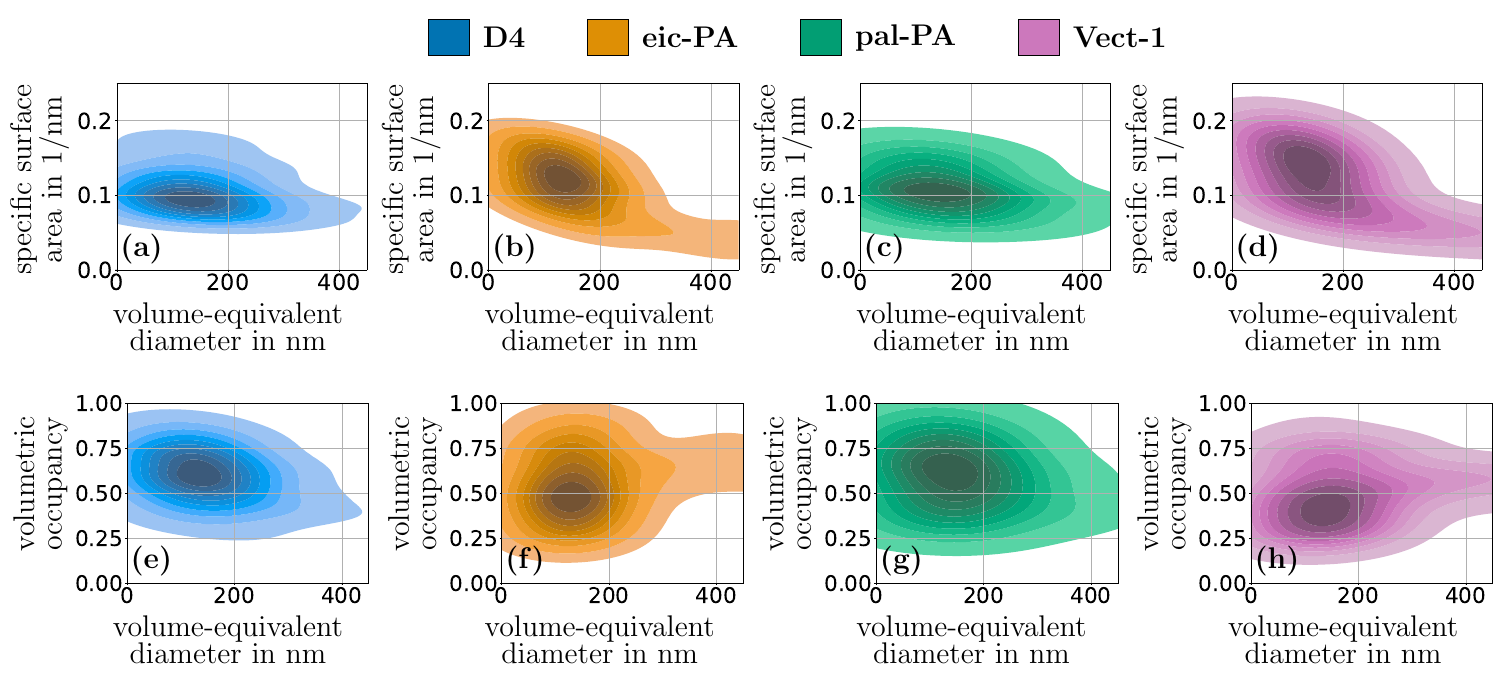}
    \caption{The bivariate probability densities between volume-equivalent diameter and specific surface area ((a)-(d)) as well as volumetric occupancy ((e)-(h)) confirm the grouping of D4 and pal-PA as well as eic-PA and Vect-1 as introduced in Section~\ref{sec:Discussion}.}
    \label{fig:biv_kde_vs_desc}
\end{figure}

Since the volume-equivalent diameter $\ved(\peptide)$ of an aggregate $\peptide\in\peptideSet$, as a descriptor of size, is biased due to the size of the observation window, it is not discussed in the manuscript, see Section~\ref{ssec:edgeEffects} for details.
Although the volume-equivalent diameter is not quantitatively reliable in this setup, it can be utilized to obtain qualitative insight into the behavior of the different peptides.
Recall Section~\ref{sec:Discussion}, where it was observed that the four peptides can be grouped based on their morphological characteristics.
D4 and pal-PA, as well as eic-PA and Vect-1, show a similar correlation structure between volume-equivalent diameter and specific surface area (Figures~\ref{fig:biv_kde_vs_desc}(a)-(d)), as well as volumetric occupancy (Figures~\ref{fig:biv_kde_vs_desc}(e)-(h)).
This similarity in the bivariate probability densities supports the previously introduced grouping.

\end{document}